\documentstyle[12pt]{article}                 
\begin{document}                                                                
\newcommand {\beq}{\begin{equation}}
\newcommand {\eeq}{\end{equation}}
\newcommand {\barra}{|}
\newcommand {\spin}{\mbox{\small$\frac{\pi}{2}$}}
\newcommand {\Tr}{\mathop{{\rm Tr}}}
\newcommand{\xs}{\mbox{$x(\sigma)$ }}
\newcommand{\vs}{\vbox{\vskip 1cm plus .3cm minus .3cm}}
\newcommand{\be}{\begin{equation}}
\newcommand{\ee}{\end{equation}}
\newcommand{\bea}{\begin{eqnarray}}
\newcommand{\eea}{\end{eqnarray}}
\newcommand{\bean}{\begin{eqnarray*}}
\newcommand{\eean}{\end{eqnarray*}}
\newcommand{\stackrange}[2]{{\scriptstyle #1 \atop \scriptstyle #2}}
\newcommand{\balpha}{\hbox{\boldmath$\alpha$}}
\newcommand{\brho}{\hbox{\boldmath$\rho$}}
\newcommand{\vac}{|0\rangle}
\newcounter{appendice}
\newcommand{\appendice}
{\setcounter{equation}{0}
\renewcommand{\theequation}{\Alph{appendice}.\arabic{equation}}
\addtocounter{appendice}{1}
{\bf Appendix \Alph{appendice}}}

\thispagestyle{empty}                                                          

\begin{flushright}
\end{flushright}

\bigskip                                                                        
\begin{center}                                                                  
{\large {\bf   Witten's Cubic Vertex in the Comma Theory (I)}}\\  
\bigskip                                                                       \end{center}     
\vspace{1cm}                                                            

\begin{center}

{\large A. ABDURRAHMAN}\\
abdurrahman\,@\,v2.rl.ac.uk \\
{\it Rutherford Appleton Laboratory,\\
  Chilton, Didcot, Oxon OX11 0QX, United Kingdom}\\
\vspace{.2cm}
{\large Jos\'e BORDES\footnote{Also at IFIC, Universitat de Valencia-CSIC
(Centro Mixto)}}\\
jose.m.bordes\,@\,uv.es\\
{\it Dept. Fisica Teorica, Univ. de Valencia,\\
  c. Dr. Moliner 50, E-46100 Burjassot (Valencia), Spain}                  

\end{center}                                                                    
                                                                                
\centerline{ABSTRACT}                                           
                                                                                
\begin{quote} 
In is shown explicitly that the
Witten's interaction 3-vertex is a solution to the  comma overlap
equations; hence  establishing the equivalence between the conventional and the
``comma" formulation of interacting string theory at the level of vertices. 
\end{quote}                                                                     
\vspace{1cm}

\noindent
Pacs: 11.10.-z,11.25.Sq

\newpage                                                                        
\section{Introduction}

In the comma approach to string field theory, the theory of open-bosonic-string
is formulated as a generalized Yang-Mills theory \cite{ChanHM}. In the comma
approach instead of choosing the bosonic coordinates $X_{0}$ of the center of
mass and the coordinates $X_{n}$ for all other points relative to it as
variables to describe the string, one chooses as an alternative the coordinates 

\begin{eqnarray}
\chi^{r} (\sigma,\tau) = 
\left\{
\begin{array}{ll}
X(\sigma,\tau),\,\,\,\,\, if\,\, r=1 \,,\\
X(\pi - \sigma,\tau),\,\,\,\,\, if\,\, r=2\,,\\
\sigma \in [0,\pi/2)\,,
\end{array}
\right.
\nonumber
\end{eqnarray}
where $r=1,2$ refers to the left ($L$), right ($R$) parts of the string and 
$\chi^{1,2}$ are restricted by 

$$
\lim_{\sigma\rightarrow \pi/2}\chi^{1} (\sigma,\tau) = 
\lim_{\sigma\rightarrow \pi/2}\chi^{2} (\sigma,\tau). 
$$
Now the midpoint\footnote{Another approach in which the midpoint plays a
central role has been discussed in reference
 \cite{MIDP}, although in a different
context from references \cite{Bordes,ABDU2}.}  varies freely  while the 
variation of the neighboring points on the string is restricted.

In \cite{Bordes,ABDU2} a  Fock space representation of the comma 
theory was developed. There one introduces comma operators,
$b_{k}^{L}$, $b_{k}^{R}$ (corresponding to each half of the string) satisfying
the commutation relations\footnote{The space-time
indices will be suppressed for the rest of the paper.}
$$
\left[b_{k}^{r,\mu}, b_{l}^{s,\nu}\right] = {\delta}^{rs} {\delta}_{n+m,0}
{\eta}^{\mu,\nu}, 
$$
analogous to the commutation relations of $a_{n}^{\mu}$ of the full string. The
Fock space corresponding, to each half (comma) of the string was introduced in
\cite{Bordes,ABDU2} leading to a basis for the Fock space of our
string in which the overall motion of the string is unrestricted but
neighboring points on the string are not permitted to wander far from each
other. The change of basis from the full string to the comma theory is not a
simple one . The explicit transformation relating both of them has been
developed in \cite{Bordes,ABDU2}. The commonly used Witten's string field
theory \cite{Witten} (WSFT) has of
course the merit of being simple for constructing solutions of the free-string
equation of motion. On the other hand,
the comma theory has the advantage of being more transparent
in questions connected with gauge invariance. In \cite{ABDU2} we have
addressed the question of equivalence between the conventional and the comma
formulations of interacting string field theory, also we examined
the relationship of the interacting vertex in both theories. In fact we were 
able to show  directly that the identity vertex ($\int$) , the $2-Vertex$ 
and the $4-Vertex$ in WSFT constructed in \cite{Gross1,CST} 
are solutions to the
overlap equations in the comma theory, hence establishing the equivalence (at
least) at the level of the vertex for $N=1,2,4$. However, the assumption that 
the $cubic$ interaction vertex (which is of central importance) 
also solves the comma
overlaps was not verified,
and we accepted that $V_{3}$ = $I\, V_{4}$ (true in Witten's 
theory) continue to hold in the comma theory. This serious shortcoming is now
removed by giving a direct proof that the $cubic$ vertex in WSFT is indeed a
solution of the comma overlaps.

The results presented  here are 
obtained in the same way as in \cite{ABDU2}; the details, however, are 
rather more involved, as one might have anticipated. Although several 
computational details are relegated to the Appendices, this paper is not meant to 
be self-contained in the sense that we relay on \cite{ABDU2} for notation
and indeed for many other details only alluded to below. For readers interested
in the formulation of the comma theory, a cursory reading of reference 
\cite{ChanHM} should suffice. Other subtleties related to the 
anomaly
cancellation for the identity vertex, the two vertex, the $3-vertex$ and the
$4-vertex$ were discussed  in \cite{ABDU2}.

\section{Comma Vertices}

In the comma formulation of string field theory the elements of the 
theory are defined by $\delta$-function 
type  overlaps. The $N$ interaction vertex is given by
\bean
&& V[\chi^{r}_{1},\chi^{r}_{2},...,\chi^{r}_{N},\varphi^{r}] =
\nonumber \\
&&
e^{iQ^{\varphi}(\phi/2)}
\prod_{i=1}^{N}\prod_{\sigma=0}^{\pi/2}\delta(\chi^{1}_{i}(\sigma)-  
\chi^{2}_{i-1}(\sigma)) 
\delta(\varphi^{1}_{i}(\sigma)-  
\varphi^{2}_{i-1}(\sigma)) 
\, .
\eean  
The index $i$ refers to the $i$th string (it is understood that $i=0$ and 
$N$ are identified) and $1$, $2$ refer to the left and right parts of the
string respectively\footnote{To avoid confusion we may use the alternative label
$L(R)$  to refer to the $left\,(right)$ parts of the string respectively.}. 
As it is seen, the ghost $\delta$-function has the same 
structure as the coordinates one and $Q^{\varphi}$ is the ghost number 
insertion. The formulation of the elements of the comma theory in the oscillator
Hilbert space is given in references \cite{Bordes,ABDU2} where they are 
represented as exponentials of quadratic forms. 
Here we are interested in the $cubic$ interaction vertex 
is given by
$$
|V_{3}^{C}> = e^{-\sum_{i=1}^{3} {b^{L}}^{\dag}_{i\,n} {b^{R}}^{\dag}_{i-1\,n}}
\prod_{i=1}^{3} |0>^{L}_{i}|0>^{R}_{i}.
$$
This vertex has the advantage of being more transparent and it is trivial to
generalize to an arbitrary number of strings which is not true in the case of
the standard formulation of string field theory 
\cite{Gross1,CST}. In fact 
the generalization of the above vertex is simply
$$
|V_{N}^{C}> = e^{-\sum_{i=1}^{N} {b^{L}}^{\dag}_{i\,n} {b^{R}}^{\dag}_{i-1\,n}}
\prod_{i=1}^{N} |0>^{L}_{i}|0>^{R}_{i}.
$$
In the oscillator Hilbert space of the comma 
theory \cite{ABDU2}, the $\delta$ functions, for the 
coordinates (the ghost degrees 
of freedom, in the bosonized representation, have the same 
structure apart from some mid-point insertions which will be addressed later)
translate into operator overlap equations, namely 
$$
\left[\chi^{L}_{i}(\sigma) - \chi^{R}_{i-1}(\sigma)\right] |V_{N}> = 0\, ,
\sigma\in[0,\pi/2) \, ,
$$
and $i=1,2,..,N$. In addition, conservation of momentum
requires
$$
\left[{\wp}^{L}_{j}(\sigma) + {\wp}^{R}_{j-1}(\sigma)\right] |V_{N}> = 0
\, .
$$
These are now the overlaps  defining equations for the comma vertices. To
simplify the calculation we introduce a new set of coordinates. Following 
Gross and Jevicki \cite{Gross1} we define
\be
{\cal Q}^{r}_{k}(\sigma) =
\frac{1}{\sqrt{N}}\sum_{l=1}^{N}\chi^{r}_{l}(\sigma) e^{\frac{2\pi i l k}{N}},\,
\,r=1,2,
\label{CV6}
\ee  
and similar ones for the momenta. The corresponding creation and annihilation
operators introduced in \cite{ABDU2} are 
defined similarly and (for $N\geq3$) satisfy the commutation relations
$$
\left[{\cal B}^{r}_{n},\overline{{\cal B}^{s}}_{-m}\right]= \delta^{rs}
\delta_{nm},\,\,\,\,\,\,\,\,\left[{\cal B}^{r}_{n},{\cal B}^{s}_{-m}\right]= 0
\, .
$$  
The advantage of this new set of variables 
is that it leads to the separation of degrees of freedom in the
overlap equations. In the complex coordinates (\ref{CV6}), the overlap 
conditions  $\chi^{L}_{j}(\sigma) = \chi^{R}_{j-1}(\sigma)$ for the three 
string vertex, read
\begin{eqnarray}                   
{\cal Q}^{L}(\sigma)  & = &   e^{2\pi i/3} {\cal Q}^{R}(\sigma)
\,\,\, ,\sigma\in
[0,\pi/2) \, ,
\nonumber \\
{\cal Q}^{L}_{3}(\sigma)  & = &  {\cal Q}^{R}_{3}(\sigma)\,\,\, ,\sigma\in
[0,\pi/2) \, ,
\label{CV8a}
\end{eqnarray}
where ${\cal Q}^{r}(\sigma) \equiv  {\cal Q}^{r}_{1}(\sigma) =
(\overline{{\cal Q}^{r}_{2}(\sigma)})$. For the complex momenta the overlap 
conditions ${\wp}^{L}_{j}(\sigma) = - {\wp}^{R}_{j-1}(\sigma)$ translate into 
\begin{eqnarray}                          
{\cal P}^{L}(\sigma) & = & - e^{2\pi i/3} {\cal P}^{R}(\sigma)\,\,\, ,\sigma\in
[0,\pi/2) \, ,
\nonumber \\
{\cal P}^{L}_{3}(\sigma) & = & - {\cal P}^{R}_{3}(\sigma)\,\,\, ,\sigma\in
[0,\pi/2) \, ,
\label{CV9a}
\end{eqnarray}
where ${\cal P}^{r}(\sigma) \equiv  {\cal P}^{r}_{1}(\sigma) =
(\overline{{\cal P}^{r}_{2}(\sigma)})$. The overlap conditions on 
${\cal Q}^{r}(\sigma)$ and
${\cal P}^{r}(\sigma)$ determine the form of the comma $3-Vertex$. The
interaction vertex, in the comma theory is          
$$                                                            
|V_{3}^{C}> = exp \left( -\frac{1}{2} ({\cal B}^{\dag}_{3}| 
{\cal I}| {\cal B}^{\dag}_{3}) -  ({\cal B}^{\dag}| {\cal H}| 
\overline{{\cal B}^{\dag}}) \right)\prod_{i=1}^{3}|0>^{1}_{i}
|0>^{2}_{i}\,,            
$$                                              
with concrete matrices ${\cal I}$ and ${\cal H}$ (see reference \cite{ABDU2}). 

Now our task is to show that Witten's $cubic$ vertex constructed in  
\cite{Gross1,CST} is
a solution of the above overlap equations. At this point it is worth mentioning
that it is not a trivial matter to show this, since one encounters various 
double infinite sums (the second coming from integrating $\sigma$
over the range $[0,\pi/2)$ in formulating the comma theory).  
 Here  the double infinite sums may not converge
absolutely and the convergence may depend on the order of the sums (so the
expressions may be ambiguous). In fact when performing these sums using the
theory of contour integration one ends up with many divergent pieces of
different orders. The amazing thing is that
all these infinities of different magnitudes conspire among them selves to leave
us with a finite number.
The case of the full string \cite{Gross1,CST} is different, 
the expression for the vertices involve
absolutely convergent sums. This ambiguity is not an accident, we have seen in
\cite{Bordes,ABDU2} that Witten's theory can be viewed as an infinite 
dimensional local matrix
algebra where the star product $``*"$ becomes matrix multiplication over
infinite dimensional matrices that does not conserve associativity. In the
standard formulation of string field theory, the Witten's
$cubic$ vertex in oscillator basis (see reference \cite{Gross1} for details
and notation) is given by
$$
|V^{W}_{3}> = exp\left(
-\frac{1}{2} ({\cal A_{3}}^{\dag}|C| {\cal A_{3}}^{\dag}) 
 -  ({\cal A}^{\dag}|U| {\overline{{\cal A}^\dag}})\right) 
\prod_{i=1}^{3}|0>_{i} \,\, ,
$$  
where $C_{nm} = (-)^{n}\delta_{nm}$ and $U_{nm}$ is an infinite dimensional
matrix  constructed in  \cite{Gross1}. The explicit form of the matrix $U$ is
given in Appendix A.  With the help of the change of representation formulas
(see reference \cite{ABDU2}) we are able to show that the
comma overlaps are satisfied  by the cubic vertex, i.e. we will     
show that the comma overlaps continue to hold in the Oscillator Hilbert
space of the Witten's cubic vertex.

\section{Proving the Overlap Equations}

\noindent
\underline{Coordinate Overlaps}.\,\,
We first note that the second equation of the coordinates overlaps 
(\ref{CV8a}), is the same as the overlap equation for the
identity vertex and therefore the proof  follows from the form of the
vertex \cite{ABDU2}. 
Hence we are only left with the first equation in
(\ref{CV8a}) to verify. The overlap conditions on 
${\cal Q}^{r}(\sigma)$, equation
 (\ref{CV8a}),  imply that their Fourier components 
satisfy 
$$
\left[ 
{\cal Q}^{L}_{2n}- e^{2\pi i/3} {\cal Q}^{R}_{2n} 
\right]|V_{3}> = 0   \,\, .
n\geq 0
$$      
Using the change of representation (see reference \cite{ABDU2}) the above equation,
in the full string oscillator Hilbert space reads
\begin{eqnarray}
\left[ -i\sqrt{3} {\cal Q}_{0} + \frac{2\sqrt 2}{\pi}
\sum_{n=0}^{\infty}\frac{(-)^{n}}{2n+1} {\cal Q}_{2n+1} \right]
|V^{W}_3>  & = &  0 \, ,
\nonumber \\
\left[ -i\sqrt{3} {\cal Q}_{2n} - 2
\sum_{m=0}^{\infty} B_{2n\, 2m+1} {\cal Q}_{2m+1} \right]
|V^{W}_3> & = &  0 \, .
\label{CV13a}
\end{eqnarray} 
Commuting the
annihilation operators in (\ref{CV13a}) through the creation operators in
$|V^{W}_{3}>$ yields a sum of creation operators acting on $|V^{W}_{3}>$, 
hence                       
$$
-\sum_{m=0}^{\infty}\left[.........\right] {\cal A}^{\dag}_{m} 
|V^{W}_{3}> \, ,
$$ 
where the expression in the square bracket is given by 
\begin{equation}
\frac{\sqrt{3}}{2} ( U_{m\,\,0} + {\delta}_{m\,\,0}) +
\frac{2i}{\pi}\sum_{n=0}^{\infty}\frac{(-)^n}{(2n+1)^{3/2}} 
(U_{m\,\,2n+1}+ {\delta}_{m\,\,2n+1}) \,\, . 
\label{CV15}
\end{equation}   
Since the states $A^{\dag}_{m}|V^{W}_{3}>$ are linearly independent, the
expression in (\ref{CV15}) must vanish for all values of $m$.
Now there are three cases to consider $m=0,\, 2k\geq 2$ and $2k+1\geq 1$. For 
$m=0$, equation (\ref{CV15}) reduces to   
\begin{equation}
\left[ \frac{\sqrt{3}}{2} ( U_{0\,\,0} + 1) +
\frac{2i}{\pi}\sum_{n=0}^{\infty}\frac{(-)^k}{(2n+1)^{3/2}} 
U_{0\,\,2n+1} \right]  \, .
\label{CV16}
\end{equation}   
Using the expression for $U_{nm}$ in reference \cite{Gross1} (see also Appendix A), 
equation (\ref{CV16}) becomes                          
\begin{equation}
\left[ \frac{\sqrt{3}}{2} ( U_{0\,\,0} + 1) -
\frac{2}{\pi}( 1 - U_{0\,\,0} )
\sum_{n=0}^{\infty}\frac{a_n}{(2n+1)^2} \right] ,  
\label{CV17}
\end{equation}                                                 
where $a_n$ are the coefficients obtained in the Taylor's expansion 
of the function 
$((1+x)/(1-x))^{1/3}$. The sum in the above expression can be found explicitly
using the contour integral representation of the same coefficients: 
$$
a_n = \frac{1}{2\pi i} {\oint}_{0} dz
\frac{1}{z^{n+1}}\left(\frac{1+z}{1-z}\right)^{1/3} .
$$                             

Now, to  perform the sum we notice 
that in the complex plane we have the cuts along the
lines $(1,\infty)$ and $(-1,\infty)$. Deforming the contours and picking the
integrals along the cuts; then commuting the sum and the integral one sees that
the series inside the integral is easily converted to another integral over
a logarithmic function in  two integration variables (say $x,y$). Performing 
the change of variable $t=\frac{x+1}{x-1}$ followed by the change of variables 
$x = 1/t$ the whole expression reduces to various forms of special functions
which are easily read from standard tables of integrals.
Hence one obtains the following value
$$
\sum_{n=0}^{\infty}\frac{a_{2n+1}}{(2n+1)^2}=\frac{\sqrt{3}}{4}\pi (3\,ln3 
- 4\,ln2) = \frac{\sqrt{3}\pi}{4}\frac{1+U_{0\,0}}{1-U_{0\,0}} \, .
$$
This proves that  equation (\ref{CV17}) is identically
zero for $m=0$ . The next case to consider is $m=2k$. In this case,
equation (\ref{CV15}) becomes 
$$
\left[ \frac{\sqrt{3}}{2}  U_{2k\,\,0} +
\frac{2i}{\pi}\sum_{n=0}^{\infty}\frac{(-)^n}{(2n+1)^{3/2}} U_{2k\,\,2n+1} 
\right] \, . 
$$                                                   
Using the explicit values of the matrix $U$, the above expression
reduces to
                            
$$
-\frac{\sqrt{3}}{2} (1- U_{0\,0})\frac{A_{2k}}{\sqrt{2k}}
+\frac{2i}{\pi}
\sum_{n=0}^{\infty}\frac{(-)^n}{(2n+1)^{3/2}}\left[\frac{-i}{2}
(2k)^{1/2}(2n+1)^{1/2}
\right.
$$

$$
\left\{\frac{A_{2k} B_{2n+1} - B_{2k}
A_{2n+1}}{(2k)+(2n+1)}  
+  \frac{A_{2k} B_{2n+1} + B_{2k}
A_{2n+1}}{(2k) - (2n+1)}\right\} + 
$$

\begin{equation}
\left. + i (1- U_{0\,0}) \frac{A_{2k}
A_{2n+1}}{(2k)^{1/2}(2n+1)^{1/2}}\right] \, ,
\label{CV21}
\end{equation}
where $A_n$ are the coefficients given by
$$
\left(\frac{1+ix}{1-ix}\right)^{1/3}= \sum_{n=0}^{\infty} A_{2n} x^{2n}
+ i\sum_{n=0}^{\infty} A_{2n+1}  x^{2n+1}
$$                             
and the corresponding coefficients $B_{n}$, are obtained in the expansion of
$((1+ix)/(1-ix))^{2/3}$.  It is worth mentioning that there is only a sign
deference between them and the analogous coefficients $a_{n}$ and $b_{n}$ 
in the expansion of the functions $((1+x)/(1-x))^{1/3}$ and 
$((1+x)/(1-x))^{2/3}$
\begin{eqnarray}
A_n & = &  (-)^{n/2} a_n\,,\,\,\,\,\,\,n=2k \, ,
\nonumber \\
A_n & = &  (-)^{(n-/2)} a_n\,,\,\,\,\,\,\,n=2k+1 \, ,
\nonumber
\end{eqnarray}
and likewise for the B's. All the sums needed in the above expression are 
worked out in the Appendices.
Putting the explicit values for all the sums we 
see that (\ref{CV21}) is
identically zero\footnote{We are skipping the details so far since 
some of the sums needed to finish the proof at later stages are more 
involved and will be given in detail.}. 
The last case to consider is
$m=2k+1\geq1$. In this case  equation (\ref{CV15}) reduces to 
 \be                   
\left[ \frac{\sqrt{3}}{2} U_{2k+1\,\,0} +        
\frac{2i}{\pi}\sum_{k=0}^{\infty}\frac{(-)^n}{(2k+1)^{3/2}} (U_{2k+1\,\,2n+1} 
+ {\delta}_{k\,n})\right],
\label{CV24}
\ee
where the matrix $U$ has the property 
$$
U_{n\, m} = (-)^{n+m} U_{m\, n} \, ,
$$
and is explicitly given by:
\begin{eqnarray}
U_{2k+1\,2n+1}& = & \frac{1}{2}\left(\frac{2k+1}{2}\right)^{1/2}  
\left(\frac{2n+1}{2}\right)^{1/2} 
\left( E(U'+{\overline{U}}')E \right)_{2k+1\,\,2n+1}
\nonumber \\
&& - \frac{U_{2k+1\,\,0} U_{0\,\,2n+1}}{1-U_{0\,\,0}} \, .
\label{CV25}
\end{eqnarray}

In this case it is a tedious job to perform the sums in (\ref{CV24}) 
due to the complicated form of the matrix $(E(U'+{\overline{U}}')E)$. 
This complication mainly 
arises when $n=k$ since in this case the expression for the 
matrix $(E(U'+{\overline{U}}')E)$ becomes ill defined and one must consider a 
limiting procedure ($n\rightarrow k$). The explicit expression of 
the matrix 
$(E(U'+{\overline{U}}')E)$ for the diagonal and off diagonal elements  is give
in reference \cite{Gross1}; however we shall rephrase them in a 
compact form (see Appendix A).  First lets consider the sum over the
matrix $U$ in (\ref{CV24}). It is clear from (\ref{CV25}) that this sum
consists of two terms, one over the nonzero elements and one over the zero
elements, we call these two terms $\Sigma_1$ and  $\Sigma_2$ respectively.
Using the explicit value of $(E(U'+{\overline{U}}')E)_{2k+1\,2n+1}$, $\Sigma_1$
reduces to
 
$$
\Sigma_1 = \frac{1}{\sqrt{2}}
\left(\frac{2k+1}{2}\right)^{1/2}\frac{(-)^k}{(2k+1)} \left[2 a_{2k+1}
\Sigma_{0}^{b} - \frac{2}{(2k+1)}\right.
$$

\be
\left. +\frac{1}{(2k+1)} \sum_{n=0}^{\infty} \frac{b_{1} a_{2n+1} - a_{1} 
b_{2n+1}}{2n} \right] +\frac{1}{2} \frac{(-)^k}{(2k+1)^{3/2}}\, ,
\label{Sigma1}
\ee
where we have made use of the identities (see Appendix B)                      \begin{eqnarray}
\sum_{n=0}^{\infty} \frac{a_{2k+1}b_{2n+1} + b_{2k+1}a_{2n+1}}{(2k+1)+(2n+1)} 
& = & 
\frac{2}{2k+1}
\nonumber \\                                                                   
\sum_{n=0}^{\infty} \frac{a_{2k+1}b_{2n+1} - b_{2k+1}a_{2n+1}}{(2k+1)-(2n+1)} 
& = &
\frac{1}{2k+1} \sum_{n=0}^{\infty} \frac{b_{1} a_{2n+1} - 
a_{1} b_{2n+1}}{(2n+1)-1} \, .
\label{CV26}
\end{eqnarray}                                                                   
The sum in (\ref{Sigma1}) is potentially divergent and must be treated 
carefully.
The term for $n=0$ should be understood as the limit of $n\rightarrow 0$ 
which follows from our definition of the diagonal elements for the matrix 
$(E(U'+{\overline{U}}')E)$ (see Appendix A). Hence one writes
\be
\sum_{n=0}^{\infty} \frac{b_{1} a_{2n+1} - a_{1} b_{2n+1}}{2n} = 
\lim_{n\rightarrow 0} \frac{b_{1} a_{2n+1} - a_{1} b_{2n+1}}{2n} + 
 \sum_{n=1}^{\infty} \frac{b_{1} a_{2n+1} - a_{1} b_{2n+1}}{2n} \, ,
\label{CV28}
\end{equation}                                                         
the numerical values of the limit and the sum in the above expression are  
$-1/3$ and $-2/3$ respectively (see Appendices A and B). Using this identity
and the numerical value
$\Sigma_{0}^{b}= \frac{1}{2}\pi\sqrt{3}$, $\Sigma_{1}$ reduces to
$$
\Sigma_{1} = \frac{1}{2}\pi\sqrt{3} \frac{(-)^k}{(2k+1)^{1/2}}a_{2k+1} -
\frac{(-)^k}{(2k+1)^{3/2}} \, .
$$                                                         
Conversely, with the values of the zero elements for the matrix $U$, 
$\Sigma_2$ is given by
$$
\Sigma_{2} = -\frac{\pi}{4}\sqrt{3}(1+U_{00})\frac{(-)^k}{\sqrt{2k+1}} 
a_{2k+1}  
$$
Now, combining $\Sigma_{1}$ and $\Sigma_{2}$ we obtain
\begin{eqnarray} 
\sum_{n=0}^{\infty} \frac{(-)^{n}}{(2n+1)^{3/2}} U_{2k+1\,2n+1} & = &
\nonumber \\
\frac{1}{2}
\pi\sqrt{3} \frac{(-)^{k}}{(2k+1)^{1/2}}a_{2k+1} - \frac{(-)^{k}}{(2k+1)^{3/2}}
& -& \frac{\pi}{4}\sqrt{3}U_{00}\frac{(-)^{k}}{\sqrt{2k+1}} a_{2k+1}\, .   
\nonumber \\
\label{CV31}
\end{eqnarray}              
Finally 
using the explicit value of $U_{2k+1\,0}$ and (\ref{CV31}) we see that the
expression in the square bracket in (\ref{CV24}) vanish. This completes the
proof of the first equation in (\ref{CV13a}). 

It remains to verify the second 
equation in (\ref{CV13a}). Again commuting the annihilation 
operators in the insertion through the creation part of $|V_{3}^{W}>$ we end 
up with    
$$
-\sum_{m=0}^{\infty}\left[.........\right] {\cal A}^{\dag}_{m} 
|V^{W}_{3}> \, ,
$$ 
where the expression in the squared bracket is now given by 
\be 
\frac{\sqrt{3}}{2}\left(\frac{2}{2n}\right)^{1/2} 
( U_{m\,\,2n} + {\delta}_{m\,\,2n}) -i
\sum_{k=0}^{\infty} B_{2n\,\,2k+1} \left(\frac{2}{2k+1}\right)^{1/2} 
(U_{m\,\,2k+1}+ {\delta}_{m\,\,2k+1}) \,\, .  
\label{CV33}
\ee                       
We have to show that this expression vanish for all values of $m\geq 0$.
The case for $m=0$ is straightforward to demonstrate and there is no need to
give it here. The other two
cases (i.e., $m= 2l\geq 2\,\,\,2l+1\geq 1$), however, are much harder. 
For $m=2l$ the above expression reduces to
\be 
\frac{\sqrt{3}}{2}\left(\frac{2}{2n}\right)^{1/2} 
( U_{2l\,\,2n} + {\delta}_{l\,\,n}) -i
\sum_{k=0}^{\infty} B_{2n\,\,2k+1} \left(\frac{2}{2k+1}\right)^{1/2} 
U_{2l\,\,2k+1} \,\,  .
\label{CV34}
\ee                
Substituting the explicit values for the matrices $B$ and $U$ and performing
some of the sums with the help of the following identity, which can be
verified using the results in Appendix A,
\begin{eqnarray}
\sum_{m=0}^{\infty} B_{2n\,2m+1} \frac{A(B)_{2m+1}}{2m+1  \pm 2l} & = & 
\nonumber  \\
\frac{(-)^n}{\pi}\left[\frac{1}{2n \mp 2l}
\left(\Sigma_{2n}^{a(b)} - \Sigma_{\pm 2l}^{a(b)}\right) 
\right. & + & \left. 
\frac{1}{2n \pm 2l} \left(\pm \Sigma_{\pm 2l}^{a(b)} - \Sigma_{- 2n}^{a(b)}
\right)\right],\,\,\,l\neq n  \, ,
\nonumber 
\end{eqnarray}
and using the result in the Appendices, the above expression takes the form
(for $l\neq n$) to
\begin{eqnarray}
& &
\! \! \! \! \! \! \! \! \! \! \! \!
\frac{\sqrt{3}}{2}  \left(\frac{2}{2n}\right)^{1/2} 
( U_{2l\,\,2n} + {\delta}_{l\,\,n}) +
\frac{(-)^{n}}{\pi}\left(\frac{2l}{2}\right)^{1/2} A_{2l} 
\nonumber \\              
& &
\! \! \! \! \! \! \! \! \! \! \! \!
\left[  \frac{1}{2l-2n} 
\left(\frac{3}{2}\Sigma_{2n}^{b} - \frac{3}{2}
\Sigma_{2l}^{b}\right)+  
\frac{1}{2l+2n}\left(\frac{3}{2}\Sigma_{2n}^{b}  -\frac{3}{2} \Sigma_{2l}^{b}
\right)\right]-
\nonumber \\
& &
\! \! \! \! \! \! \! \! \! \! \! \! 
-\frac{(-)^{n}}{\pi}\left(\frac{2l}{2}\right)^{1/2} B_{2l}
\left[ \frac{1}{2l-2n}
\left(\frac{3}{2}\Sigma_{2n}^{a}  -\frac{3}{2}
\Sigma_{2l}^{a} \right)-  
\frac{1}{2l+2n}\left(\frac{3}{2}\Sigma_{2n}^{a} +\frac{3}{2}\Sigma_{2l}^{a}
\right)\right] +
\nonumber \\
& &
\! \! \! \! \! \! \! \! \! \! \! \!
+\sqrt{\frac{3}{2}} (1-U_{0\,\,0}) \frac{A_{2n}}{2n}\frac{A_{2l}}{\sqrt{2l}}
\, ,
\nonumber \\
\label{CV35}
\end{eqnarray}
where
$$
\Sigma_{\pm {2n}}^{a(b)} = \sum_{m=0}^{\infty}\frac{a(b)_{2m+1}}{\pm {2n}+2m+1}
\, ,
$$
and we have also made use of the identities
\begin{eqnarray}
& &
\sum_{k=0}^{\infty} B_{2n\,\,2k+1} \frac{A_{2k+1}}{2k+1} =\frac{\sqrt{3}}{2}
\frac{A_{2n}}{2n} \, ,
\nonumber \\
& &
\Sigma_{-2n}^{a} = -\frac{1}{2}\Sigma_{2n}^{a}\,\,,
   \Sigma_{2n}^{a} = +\frac{1}{2}\Sigma_{2n}^{b} ,
\nonumber
\end{eqnarray}
which can be verified using the integral representation of the coefficients 
$a's$ and $b's$ . All other sums involved above are given in reference 
\cite{Gross1} and Appendices A and B. Recalling that
$$
U_{2l\,\,2n}= U_{2l\,\,2n}^{\prime} -\frac{U_{2l\,\,0}
U_{0\,\,2n}}{1-U_{0\,\,0}},
$$
with
$$
U_{2l\,\,2n}^{\prime} = \frac{1}{2}\sqrt{\frac{2l}{2}}\sqrt{\frac{2n}{2}}
\left( E(U'+{\overline{U}}')E \right)_{2l\,\,2n},
$$
we see, after some algebra, that the expression in (\ref{CV35}) vanish. To
complete this case we still need to consider $n = l$.  For this case the
expression in (\ref{CV15}) is
$$
\frac{\sqrt{3}}{2}\left(\frac{2}{2n}\right)^{1/2} 
( U_{2n\,\,2n} + 1) -i
\sum_{k=0}^{\infty} B_{2n\,\,2k+1} \left(\frac{2}{2k+1}\right)^{1/2} 
U_{2n\,\,2k+1} \,\,  .
$$   
This expression involves many sums  of the form $\Sigma_{\pm 2n}^{a(b)}$ and
${\tilde{\Sigma}}_{\pm 2n}^{a(b)}$. Where $\Sigma_{\pm 2n}^{a(b)}$ has been 
considered previously and
$$
{\tilde{\Sigma}}_{\pm 2n}^{a(b)}= \sum_{m=0}^{\infty}\frac{a(b)_{2m+1}}{(\pm
2n+2m+1)^{2}} \, .
$$
These sums can be found explicitly (see references \cite{Gross1}), however
in this case we need not evaluate them since the ones coming
from the expansion of the second term over $U$ in the above expression cancel
against the ones coming from the $U_{2n\,\,2n}$ in the first term. Thus if one 
expands the second term and substitute the explicit values for the $U$ matrix 
(these are given in  Appendix A); then with the help of the
identity \cite{Gross1}:
\begin{eqnarray}
\sum_{m=0}^{\infty} B_{2n\,2m+1} \frac{A(B)_{2m+1}}{\pm{2n} + (2m+1)}  = 
\frac{(-)^n}{\pi}
\left[\frac{1}{2(2n)}\left(\Sigma_{2n}^{a(b)} - \Sigma_{-2n}^{a(b)}\right) -  
{\tilde{\Sigma}}_{\pm 2n}^{a(b)}\right],
\nonumber 
\end{eqnarray}               
the above expression reduces to
$$
-\frac{\sqrt{3}}{4}\left(\frac{2}{2n}\right)^{1/2}
 a_{2n} b_{2n} + \frac{1}{2\pi}
\left(\frac{2}{2n}\right)^{1/2}
 b_{2n}\left(\Sigma_{2n}^{a}-\Sigma_{-2n}^{a}\right) \, ,
$$
which is identically zero since
$$
\Sigma_{-2n}^{a} = -\frac{1}{2}\Sigma_{2n}^{a} =
-\frac{1}{2}\sqrt{\frac{1}{3}}\pi a_{2n} \, .
$$
Hence (\ref{CV34}) vanish for all values of
$m=2l\geq 2$. It remains to consider
$m=2l+1\geq 1$. For  $m=2l+1\geq 1$,
(\ref{CV33}) reduces to
\be 
\frac{\sqrt{3}}{2}\left(\frac{2}{2n}\right)^{1/2} U_{2l+1\,\,2n}  -i
\sum_{k=0}^{\infty} B_{2n\,\,2k+1} \left(\frac{2}{2k+1}\right)^{1/2} 
(U_{2l+1\,\,2k+1}+ {\delta}_{l\,\,k}) \,\,  .
\label{CV45}
\ee    
The sum involving the $U$ matrix is very delicate when $k$ takes the value $l$.
We will see how to handle it below. Using the explicit values of $U$, we get
\begin{eqnarray}
T_1+T_2 & = & 
\sum_{k=0}^{\infty} B_{2n\,\,2k+1} \left(\frac{2}{2k+1}\right)^{1/2} 
U_{2l+1\,\,2k+1} =
\nonumber \\
& = & \sum_{k=0}^{\infty} B_{2n\,\,2k+1} 
\left(\frac{2}{2k+1}\right)^{1/2}
\left[\frac{1}{2}\left(\frac{2l+1}{2}\right)^{1/2}  
\left(\frac{2k+1}{2}\right)^{1/2}
\right.
\nonumber \\
& & \left. 
\left( E(U'+{\overline{U}}')E \right)_{2l+1\,\,2k+1}
- \frac{U_{2l+1\,\,0} U_{0\,\,2k+1}}{1-U_{0\,\,0}} \right] \, .
\nonumber
\end{eqnarray}
To evaluate the above sums we make use of the following 
identity\footnote{The
derivation of this identity is quite complicated; it makes use of Appendix A and
the  identities  already derived in Appendix B. The complication arise when $m$
takes the value $k$ since for this term the expression becomes ill defined. See
Appendix A for handling such terms.} 
\begin{eqnarray}
& & \sum_{m=0}^{\infty} B_{2n\,\,2m+1} \frac{A(B)_{2m+1}}{(2m+1) \pm (2k+1)} =
\nonumber \\
& & \frac{(-)^n}{\pi}
\left[\frac{1}{2n \mp (2k+1)} \Sigma_{2n}^{a(b)} 
- \frac{1}{2n \pm (2k+1)} \Sigma_{-2n}^{a(b)} - \left(\frac{1}{2n \mp (2k+1)} 
-
\right. \right.
\nonumber \\
& & \left. \left.
\frac{1}{2n \pm (2k+1)}\right) S_{\pm (2k+1)}^{a(b)}\right] \, ,
\label{CV47}
\end{eqnarray}
where
$$
S_{\pm (2k+1)}^{a(b)} = \sum_{n=0}^{\infty}
\frac{a(b)_{2n+1}}{(2n+1)\pm(2k+1)}\,\, ,
$$           
and the identities given in equations (\ref{CV26}) and
(\ref{CV28}).                           
Hence
\begin{eqnarray}
& & T_1 = \left(\frac{2l+1}{2}\right)^{1/2}
\left[\frac{\sqrt{3}}{2}\pi\left(a_{2l+1}b_{2n}+\frac{2n}{2l+1}b_{2l+1}a_{2n}
\right)- \frac{3}{2l+1}\right] B_{2n\,2l+1}
\nonumber \\
& & T_2=
-\left(\frac{3}{2}\right)^{1/2}\frac{(-)^{n+l}}{\pi}(1-U_{00})\frac{a_{2l+1}
a_{2n}}{(2l+1)^{1/2}(2n)} \, .
\nonumber
\end{eqnarray}                                          

The values of the matrix $U_{2l+1\,2n}$ are given in Appendix A. Putting
everything in equation (\ref{CV45}) we see that it is identically zero. This
completes the proof for the coordinate overlaps.

\underline{Momentum Overlaps} 

\noindent
The momentum overlaps in the comma theory are given by equation 
(\ref{CV9a}). The second equation is the same as the overlap equation of the
identity vertex and therefore the proof follows from the form of the vertex.
Now the overlap condition on ${\cal P}^{r}(\sigma)$, given by the
first equation in (\ref{CV9a}),
imply that its Fourier components
satisfy
\be
\left[P_{2n}^{L} + e^{2i\pi/3} P_{2n}^{R} \right] |V_{3} > = 0\,\,\,\,\,\, n 
\geq 0 \,\, .
\label{C1}
\ee
As well as their complex conjugate. The proof of the case for $n=0$  does not 
involve new identities other than 
those used to prove the coordinate part and therefore there is no need to give 
it here although we have checked that the overlap equation for the zero mode is
 indeed satisfied.  However, for $n\geq 1$ the proof requires deriving new 
identities . Using the change of representation formulas 
(see reference \cite{ABDU2}), equation (\ref{C1}) reduces for $n\geq 1$ to
\be 
\frac{1}{2} e^{i\pi/3}  \sum_{k=0}^{\infty} 
\left[..............\right] 
{\cal A}_{3}^{\dag} |V_{3}^{W}> \, ,
\label{C2}
\ee
where the expression in the square bracket is 
\begin{eqnarray}
\left(\frac{2n}{2}\right)^{1/2}\left(-U_{k\,2n} + \delta_{k\,2n}\right) 
& - & 
2i\sqrt{3}\sum_{m=0}^{\infty} B_{2n\,2m+1} \left(\frac{2m+1}{2}\right)^{1/2}
\nonumber \\
& \times & \left(U_{k\,2m+1} - \delta_{k\,2m+1}\right) \, .
\label{C3}
\end{eqnarray}
In order for (\ref{C2}) to vanish, (\ref{C3}) must vanish for all values of 
$k$ since the states ${\cal A}_{k}^{\dag} |V_{3}^{W}>$ are all 
linearly independent. 
Therefore to prove the overlap equation (\ref{C1}), it is sufficient to show 
that (\ref{C3}) is identically zero for all values of $k$. For $k=0$ , 
(\ref{C3}) is zero as can be easily seen by explicit substitution (no new 
identities are needed here). The other two case, i.e., $k=even \geq 2$ and 
$k=odd \geq 1$ need careful treatment since they involve quantities which are 
potentially divergent. Setting $k=2l$ in (\ref{C3}) we obtain
\be
-\left(\frac{2n}{2}\right)^{1/2}\left(U_{2l\,2n} - \delta_{l\,n}\right) - 
2i\sqrt{3}\sum_{m=0}^{\infty} B_{2n\,2m+1} \left(\frac{2m+1}{2}\right)^{1/2} 
U_{2l\,2m+1} \, .
\label{C4}
\ee 

First let us consider the case 
$l\neq n$ since it is the easier of the two cases to 
prove. Then if  one substitutes the explicit values 
of the matrix $U$ for $l\neq n$ and makes use of the following
identity                
\begin{eqnarray}
& & 
\sum_{m=0}^{\infty} B_{2n\,2m+1} \frac{(2m+1)}{2l  \pm (2m+1)} A(B)_{2m+1} = 
\frac{(-)^n}{\pi}\left[\mp 2\Sigma_{\pm 2l}^{a(b)} + \frac{2n}{2l-2n}\right.
\nonumber \\
& & 
\left. \left(\pm \Sigma_{\pm 2n}^{a(b)} \mp \Sigma_{\pm2l}^{a(b)}\right) + 
\frac{2n}{2l+2n} \left(\pm \Sigma_{\pm 2l}^{a(b)} - \Sigma_{\mp 2n}^{a(b)}
\right)\right],\,\,\,l\neq n \, ,
\nonumber 
\end{eqnarray}                         
(after a lengthy, otherwise a straight forward exercise)  we see that 
(\ref{C4}) is indeed zero. 

The case  $l=n$ contains potentially divergent 
terms and requires proving new identities involving the sums of the Taylor 
modes. Substituting the explicit values of the matrix $U$  in (\ref{C4}) and
making use of the identity,
\begin{eqnarray}
&& \sum_{m=0}^{\infty} B_{2n\,2m+1} \frac{(2m+1)}{2n  \pm (2m+1)} A(B)_{2m+1} 
=
\nonumber \\
&& 
=\frac{(-)^n}{\pi} \left[(2n){\tilde{\Sigma}}_{\pm 2n}^{a(b)}
\mp \frac{3}{2} \Sigma_{\pm 2n}^{a(b)} \mp \Sigma_{\mp2n}^{a(b)}\right]
\, ,
\nonumber 
\end{eqnarray}
we  get 
\begin{eqnarray}
2 \left(\frac{2n}{2}\right)^{1/2}  \left[1 + \frac{\sqrt3}{\pi}\frac{2n}{2}
\left( \frac{2}{3} b_{2n}  {\tilde{\Sigma}}_{2n}^{a} - \frac{4}{3} b_{2n} 
{\tilde{\Sigma}}_{-2n}^{a} -\frac{2}{3} a_{2n} {\tilde{\Sigma}}_{2n}^{b} - 
\frac{4}{3} a_{2n} {\tilde{\Sigma}}_{-2n}^{b} \right) \right] \, ,
\nonumber \\
\label{C6}
\end{eqnarray}
for $l=n$. The values of ${\tilde{\Sigma}}$ for $n\leq 0$ are related to the 
values of ${\tilde{\Sigma}}$ for $n\geq 0$ through the following identities 
(the proof is given in Appendix B) 
\begin{eqnarray}
& & {\tilde{\Sigma}}_{-n}^{a} - \frac{1}{2} {\tilde{\Sigma}}_{n}^{a} = 
\frac{3}{2} \Sigma_{0}^{a} S_{n}^{a} \, ,
\nonumber \\
& & {\tilde{\Sigma}}_{-n}^{b} +\frac{1}{2} {\tilde{\Sigma}}_{n}^{b} = 
\frac{1}{2} \Sigma_{0}^{b} S_{n}^{b} \, .
\nonumber
\end{eqnarray}
Hence (\ref{C6}) reduces to 
\be
2 \left(\frac{2n}{2}\right)^{1/2}  \left[1 - \frac{2n}{2}\left(  b_{2n}  
S_{2n}^{a}  +  a_{2n}  S_{2n}^{b} \right)\right] \, .
\label{C9}
\ee

Now using the identity (see Appendix B for derivation)
\be
a_{2n} S_{2n}^{b} + b_{2n} S_{2n}^{a} = \frac{2}{2n}  \,  ,
\label{C9.1}
\ee
we see that (\ref{C9}) is identically 
zero. This completes the proof for  $k=2l \geq 2$. Next we consider 
$k=odd=2l+1\geq 1$. For this case (\ref{C3}) reduces to
\begin{eqnarray}
\left(\frac{2n}{2}\right)^{1/2} U_{2l+1\,2n}  
+ 2i\sqrt{3}\sum_{m=0}^{\infty}
B_{2n\,2m+1} \left(\frac{2m+1}{2}\right)^{1/2} \left(U_{2l+1\,2m+1} - 
\delta_{l\,m}\right) \, .
\nonumber \\
\label{C10}
\end{eqnarray}

The explicit value of the matrix element $U_{2l+1\,2n}$ is given in Appendix A.
The sum over $U$ has not been considered before and therefore needs to be 
evaluated. However,   
in performing the sum over $U$ one has to be extra careful when $m$ takes the 
value $l$ since this term is potentially divergent. Consider the sum over the
matrix $U$ in (\ref{C10}) , i.e.,
\begin{eqnarray}
& &
\sum_{m=0}^{\infty} B_{2n\,2m+1} \left(\frac{2m+1}{2}\right)^{1/2} 
U_{2l+1\,2m+1}  =  \frac{1}{4} \left(\frac{2l+1}{2}\right)^{1/2} \Sigma_{I} +
\nonumber \\
& &
 +\frac{1}{2} \left(\frac{2l+1}{2}\right)^{1/2} B_{2n\,2l+1} 
-\left(\frac{1}{2}\right)^{1/2}  \frac{U_{2l+1\,0}}{1-U_{00}} \Sigma_{II}
\, ,
\nonumber
\end{eqnarray}
where
\be
\Sigma_{I} = \sum_{m=0}^{\infty} B_{2n\,2m+1} (2m+1) 
\left( E(U'+\overline{U}')E \right)_{2l+1\,2m+1} \, ,
\label{C12}
\ee
and
$$
\Sigma_{II} =  \sum_{m=0}^{\infty} B_{2n\,2m+1} (2m+1) ^{1/2} U_{0\,2m+1} \, .
$$
$\Sigma_{II} $ can be easily evaluated using the identity
\be
\sum_{m=0}^{\infty} B_{2n\,2m+1} A_{2m+1} = -\frac{1}{2\sqrt3} A_{2n} \, ,
\label{C14}
\ee
which can be checked by converting the sum into integral form and then 
evaluating it using the theory of special functions (we will not do it here 
since this is relatively easy to do). Hence 
$$
\left(\frac{1}{2}\right)^{1/2}  \frac{U_{2l+1\,0}}{1-U_{00}} \Sigma_{II} = 
- \frac{1}{4\sqrt3}(1-U_{00})\left(\frac{2}{2l+1}\right)^{1/2} A_{2l+1} A_{2n}ç
\, .
$$

Next we consider $\Sigma_{I}$. Substituting the explicit values of the matrices
$B$ and $U$ in (\ref{C12}) and then making use of the identities (\ref{CV26}), 
and (\ref{C14}) we get (skipping a rather tedious algebra)
\begin{eqnarray}
& &
\frac{1}{4} \left(\frac{2l+1}{2}\right)^{1/2} \Sigma_{I} = 
\frac{1}{2} \left(\frac{2l+1}{2}\right)^{1/2} \left\{ \frac{1}{\sqrt{3}} 
\left[ (-)^{n+l+1} b_{2l+1} a_{2n}\right. \right.
\nonumber \\
& &  \!\!\!\!\!
\left.  + \pi \left(\frac{2l+1}{2}\right) a_{2l+1} b_{2n} 
B_{2l+1\,2n}\right]
 -\left. \left[ \frac{\pi}{\sqrt3}\left(\frac{2l+1}{2} \right)b_{2l+1} a_{2n} 
-1\right] B_{2n\,2l+1} \right\} \, .
\nonumber
\end{eqnarray}
Putting everything together and making use of the fact 
$$
n\, B_{nm} = -m\, B_{mn} \, ,
$$
we see that (\ref{C10}) is indeed zero. This completes the proof of the $P$ 
overlaps.

\noindent
\underline{Ghost Overlaps}

To complete the proof we have to see if the ghost part of the Witten's
$cubic$ vertex satisfies (violates) the comma overlaps in exactly the same way 
as in the standard formulation. The ghost part of the Witten's vertex 
is given by 
$$
|V_{3}^{\phi}> = e^{\frac{3}{2}i\phi(\pi/2)}|V_{3}^{\phi,0}> \, .
$$    
Here, the vertex $|V_{3}^{\phi,0}>$ has the same form as the orbital part.
The ghost factor $e^{\frac{3}{2}i\phi(\pi/2)}$ corresponding to ghost number
$3/2$ is the right ghost number \cite{Witten}. Expanding the phase factor and
commuting the annihilation operator through the creation part of the vertex
results in doubling the creation part of the insertion. Thus one has
$$
|V_{3}^{\phi}> =                                           
exp\left(\sqrt{3}\sum_{n=1}^{\infty}\frac{(-)^n}{\sqrt{2n}} 
{\cal A}^{\dag}_{3\,\,2n}\right) |V_{3}^{\phi,0}> \, .
$$    
The quadratic part of the vertex $ |V_{3}^{\phi,0}>$ 
satisfies the comma overlap
equations since it has the same structure as the orbital part which solves the
comma overlaps as we have already seen. However, when one includes the ghost
insertion this is no longer the case. To see this one first observes that the
comma overlaps for $V_{3}$ are blind to the phase factor\footnote{The reason
for this is that the other overlaps describe different strings in the complex
space as we have already seen.} (insertion) apart from   
\begin{eqnarray}
{\cal Q}_{3\,\,2n}^{L} & = & {\cal Q}_{3\,\,2n}^{R}\,,\,\,\,\,n\geq 0,
\nonumber \\
{\cal P}_{3\,\,2n}^{L} & = & - {\cal P}_{3\,\,2n}^{R}\,,\,\,\,\,n\geq 0.
\nonumber
\end{eqnarray}

In fact the first of these equations is also blind to the insertion factor,
since it contains only odd modes in the annihilation-creation operator $A_{3}$
which clearly commute with the even modes in the phase factor. On the other
hand, the second equation contains even modes of the operator $A_{3}$ and
therefore is not satisfied by the vertex due to the insertion. To see this
notice that
\begin{eqnarray}
& & {\cal P}_{3\,\,2n}^{r} 
exp \left( \sqrt{3}\sum_{n=1}^{\infty}\frac{(-)^n}{\sqrt{2n}}
{\cal A}^{\dag}_{3\,\,2n}\right) =
\nonumber \\
& & 
exp \left(\sqrt{3}\sum_{n=1}^{\infty}\frac{(-)^n}{\sqrt{2n}}
{\cal A}^{\dag}_{3\,\,2n}\right)
\left[-\frac{\sqrt{3}}{2} \frac{(-)^n}{\sqrt{2n}} 
+ {\cal P}_{3\,\,2n}^{r} \right] \, .
\nonumber 
\end{eqnarray}
where $r=1,2$ refers to the left and right parts of the string respectively.
Thus commuting the overlaps through the insertion factor and collecting terms
we obtain
$$
exp \left(\sqrt{3}\sum_{n=1}^{\infty}\frac{(-)^n}{\sqrt{2n}} 
{\cal A}^{\dag}_{3\,\,2n}\right)\left[-\sqrt{3}\frac{(-)^n}{\sqrt{2n}} +
{\cal P}_{3\,\,2n}^{L} +{\cal P}_{3\,\,2n}^{R}\right] \, .
$$

Now it is clear that the overlaps in the square bracket are not satisfied by
the quadratic part of the ghost vertex because of the presence of a $c-number$.
This is the same violation seen in the operator formulation of Witten's string
field theory (see reference \cite{Gross1}). Therefore the comma overlaps are
satisfied (violated) by the Witten's $cubic$ vertex in exactly the same way as
in the case of the standard formulation\cite{Gross1,CST}. This
completes the proof that the Witten's $cubic$ vertex is a solution to the comma
overlaps.

\section{Conclusions}

We have demonstrated that the $cubic$ vertex in Witten's string
field theory, making use of the operator formalism as developed in 
references \cite{Gross1,CST}, solves the comma overlaps. However, 
as we have seen in \cite{ABDU2}, there are still few subtleties 
which must be understood if
one is to get greater confidence in the comma approach to string field theory.
For example, as pointed out in reference \cite{ABDU2}, while in  the full string 
formulation, the $K$ and the $BRST$ invariance require some specific ghost 
insertions at the midpoint of the string for consistency, it does not seem to be
the case in
the comma formalism. In the comma theory both the orbital and the ghost parts
of the $cubic$ vertex are invariant separately. Also in the same reference
we have
seen that the associativity anomaly in the star algebra  of the standard
formulation of string field theory disappears in the comma theory. The $Q$  or 
gauge invariance in the comma representation have only been mentioned in 
reference
\cite{ABDU2}. However, for a complete discussion, it is useful to use
a fermionic representation of the comma ghost. In the fermionic representation
it is possible to construct the analogous comma (ghost) $cubic$ vertex and 
examine other properties of standard string theory in the comma representation,
For completeness, we still have to show that the $cubic$ ghost vertex in the
fermionic representation of Witten's string field theory still solves the comma
overlaps of the ghost in the fermionic representation. This fact and
the proper treatment of the BRST operator $Q$ will be
given in a separate publication \cite{3VGHOST}.

\hfill \\

One of us (JB) is supported in part by grants CYCIT96-1718, PB97-1261 
and GV98-1-80.  He would also like to thank the Rutherford Appleton
Laboratory for hospitality.

\newpage

\appendice

\noindent
\underline{The Coefficients $a_{n}$ and $b_{n}$}.

In the first part of this
Appendix we give the properties of the coefficients in
the Taylor series expansion of the functions                               
\begin{eqnarray}
\left(\frac{1+x}{1-x}\right)^{1/3} & = &  \sum_{n=0}^{\infty} a_{n} x^{n},
\nonumber \\
\left(\frac{1+x}{1-x}\right)^{2/3} & = &  \sum_{n=0}^{\infty} b_{n} x^{n}.
\nonumber
\end{eqnarray}                                         
They are instrumental in the proof of the overlap equation. Most of the results
listed here are derived in reference \cite{Gross1}, or follow from results given
there. The integral form of the coefficients $a_{n}$  is given by
\be                                  
a_n = \frac{1}{2\pi i} {\oint}_{0} dz
\frac{1}{z^{n+1}}\left(\frac{1+z}{1-z}\right)^{1/3},
\label{TS3}
\ee
and likewise for the $b_{n}$ with $1/3 \rightarrow 2/3$. This form can be
utilized to derive the various recursion relations satisfied by the Taylor
modes. Integrating (\ref{TS3}) by parts leads to
$$
(n+1)a_{n+1} - \frac{2}{3} a_{n} - (n-1) a_{n-1} =0,
$$  
likewise we get for $b_{n}$      
$$   
(n+1)b_{n+1} - \frac{4}{3} b_{n} - (n-1) b_{n-1} =0.
$$       
Making use of the same integral representation, one could derive the 
cross-recursion relations
\begin{eqnarray}
\frac{4}{3} a_{n} & = & (-)^{n}\left[(n+1)b_{n+1} - 
2n b_{n} +(n-1) b_{n-1}\right],
\nonumber \\
\frac{2}{3} b_{n} & = & (-)^{n}\left[(n+1)a_{n+1} - 2n a_{n} +(n-1) 
a_{n-1}\right].
\nonumber
\end{eqnarray}        
In the text we meet various sums involving these coefficients. The primary ones
being 
\begin{eqnarray}
\Sigma_{n}^{a} \equiv \sum_{n+m=odd} \frac{a_{n}}{n+m},
\nonumber \\
S_{n}^{a} \equiv \sum_{n+m=even} \frac{a_{n}}{n+m},
\nonumber
\end{eqnarray}                          
and analogously for $a \rightarrow b$. All these sums are absolutely
convergent. All the sums given above have been evaluated in reference
 \cite{Gross1}.
Here we merely quote the results. For the sums labeled by $\Sigma$ the results
are
\begin{eqnarray}                                                    
\Sigma_{0}^{a}=\frac{1}{2}\sqrt{\frac{1}{3}}\pi,\,\,\,\,\,
\Sigma_{n}^{a}=\sqrt{\frac{1}{3}}\pi a_{n},\,\,\,\,\,
\Sigma_{0}^{a}=-\frac{1}{2} \Sigma_{n}^{a},
\nonumber \\                                                      
\Sigma_{0}^{b}=\frac{1}{2}\pi \sqrt{3},\,\,\,\,\,
\Sigma_{n}^{b}=\sqrt{\frac{1}{3}}\pi b_{n},\,\,\,\,\,
\Sigma_{0}^{b}=\frac{1}{2} \Sigma_{n}^{b},
\nonumber
\end{eqnarray} 
where $n$ is a positive integer. The results for the sums labeled by $S$ are
are given by
\begin{eqnarray}
S_{n}^{a}=\left(\frac{3}{2}- ln2\right) a_{n} +\frac{3}{2}\sum_{k=0}^{n-1} 
(-)^{k} \frac{a_{n} a_{n-k-1}}{k+1},
\nonumber \\ 
S_{n}^{b} =\left(\frac{3}{4}+ ln2\right) b_{n} +\frac{3}{4}\sum_{k=0}^{n-1} 
(-)^{k} \frac{b_{n} b_{n-k-1}}{k+1},
\nonumber
\end{eqnarray} 
for $n>1$. For $n=1$, the results are given by the same expressions without the
summations over $k$. The $S_{n}^{a}$ and $S_{n}^{b}$ satisfy the same recursion
relations as the $a_{n}$ and $b_{n}$ respectively
\begin{eqnarray}
(n+1)S_{n+1}^{a} - \frac{2}{3} S_{n}^{a} - (n-1) S_{n-1}^{a} =0 \, ,
\nonumber \\                               
(n+1)S_{n+1}^{b} - \frac{4}{3} S_{n}^{b} - (n-1) S_{n-1}^{b} =0 \, ,
\label{TS16}
\end{eqnarray} 
for $n>1$.  Another sum involving the Taylor modes which appears in the text is
$$
{\tilde{\Sigma}}_{a} \equiv \sum_{n+m=even} \frac{a_{n}}{(n+m)^{2}} \, ,
$$  
and likewise for $b$ with ($a\rightarrow b$). The values of these sums are
given in reference
 \cite{Gross1} and we shall not reproduce them here. However, it
is important to notice that these sums satisfy the following 
recursion relations
\begin{eqnarray}
(n+1) {\tilde{\Sigma}}_{n+1}^{a} = \frac{2}{3} {\tilde{\Sigma}}_{n}^{a}
+(n-1) {\tilde{\Sigma}}_{n-1}^{a} + \Sigma_{n+1}^{a} -\Sigma_{n-1}^{a},
\nonumber \\                                                     
(n+1) {\tilde{\Sigma}}_{n+1}^{b} = \frac{4}{3} {\tilde{\Sigma}}_{n}^{b}
+(n-1) {\tilde{\Sigma}}_{n-1}^{b} + \Sigma_{n+1}^{b} -\Sigma_{n-1}^{b},
\label{TS18}
\end{eqnarray} 

\noindent
\underline{The Matrix Elements $U_{nm}$}.

We shall first give the explicit 
values of the matrix elements
$U_{n\,\,m}$ appearing in Witten's vertex. They have been derived
explicitly in reference \cite{Gross1}. 
The matrices $U$ and $U'$ are related by
$$
U_{n\,\,m}= U_{n\,\,m}^{\prime} -\frac{U_{n\,\,0}
U_{0\,\,m}}{1-U_{0\,\,0}},
$$ 
where
$$
U_{n\,\,m}^{\prime} = \frac{1}{2}\sqrt{\frac{n}{2}}\sqrt{\frac{m}{2}}
\left( E(U'+{\overline{U}}')E \right)_{n\,\,m}
$$
and                                                              
\be
\left( E(U'+{\overline{U}}')E \right)_{n\,\,m} =
2(-)^{n+1}\left(\frac{A_{n}B_{m}+B_{n}A_{m}}{n+m} + 
\frac{A_{n}B_{m}-B_{n}A_{m}}{n-m}\right) \, ,
\label{A3}
\ee                                                              
for $n+m=even$ and $n,m\geq 1$\,,                                
\be
\left( E(U'-{\overline{U}}')E \right)_{n\,\,m} =
-2i\left(\frac{A_{n}B_{m}-B_{n}A_{m}}{n+m} + 
\frac{A_{n}B_{m}+B_{n}A_{m}}{n-m}\right) \, ,
\label{A4}
\ee  
for $n+m=odd$ and $n,m\geq1$. For the zero mode components
\begin{eqnarray}
\left( E(U'+{\overline{U}}')E \right)_{0\,\,m} & = &
-2\sqrt{2}\left(\frac{A_{m}}{m}\right)
\,,\,\,\,\,\,\,\,\,m=2k \,,
\nonumber \\  
\left( E(U'-{\overline{U}}')E \right)_{0\,\,m} & = &
i2\sqrt{2}\left(\frac{A_{m}}{m}\right)
\,,\,\,\,\,\,\,\,\,m=2k+1 \, .
\label{A6}
\end{eqnarray}  
The diagonal elements, $(E(U'-{\overline{U}}')E)_{n\,\,n}=0$ while
$(E(U'+{\overline{U}}')E)_{n\,\,n}$ are given by
\be
\left( E(U'+{\overline{U}}')E \right)_{n\,\,n} =
-2\left[(-)^{n}\frac{A_{n}B_{n}+1}{n}+\Delta_{n}\right] \, ,
\label{A7}
\ee                                            
where 
$$
\Delta_{n=2k}=\frac{1}{n} -\frac{2}{\sqrt{3}}\frac{(-)^{n/2}}{\pi}\left[
\left({\tilde{\Sigma}}^{a}_{n} + {\tilde{\Sigma}}^{a}_{-n}\right) B_{n}  -
\left({\tilde{\Sigma}}^{b}_{n} - 
{\tilde{\Sigma}}^{b}_{-n}\right) A_{n}\right] \,
$$
and
\be
\Delta_{n=2k+1}=\frac{3}{n} -2\sqrt{3}\frac{(-)^{(n-1)/2}}{\pi}\left[
\left({\tilde{\Sigma}}^{b}_{n} + {\tilde{\Sigma}}^{b}_{-n}\right) A_{n}  -
\left({\tilde{\Sigma}}^{a}_{n} - {\tilde{\Sigma}}^{a}_{-n}\right) B_{n}\right].
\label{A9}
\ee
The symbol ${\tilde{\Sigma}}^{a(b)}_{n}$ has been introduced before.
For completeness we also give the matrix elements of $E$. They are
$$
(E_{-1})_{n\,\,m}=\sqrt{\frac{n}{2}}\delta_{n\,\,m} + \delta_{n\,\,0} 
\delta_{m\,\,0} \, .
$$

It is worth observing that expressions for the matrix elements of $U$ for the
off diagonal elements defined by (\ref{A3}) for $n=odd$, $m=odd$ and the 
diagonal 
elements defined by (\ref{A7}) for $n=odd$ can be combined into a single 
expression. This observation
is instrumental in making the evaluation of the sums much easier. To derive a
single expression for $U$ we need the  value of $\Delta_{1}$.
Setting $n=1$ in (\ref{A9}) gives
$$
\Delta_{1}=3 -\frac{2\sqrt{3}}{\pi}\left[
\left({\tilde{\Sigma}}^{b}_{1} + {\tilde{\Sigma}}^{b}_{-1}\right) A_{1}  -
\left({\tilde{\Sigma}}^{a}_{1} - {\tilde{\Sigma}}^{a}_{-1}\right) B_{1}\right].
$$
The sums in the above expression are easily converted into integrals using the 
integral representation  of the coefficients $a_{n}$ and $b_{n}$. For example 
\begin{eqnarray}
& & 
\sum_{m=0}^{\infty}\frac{a_{2m}}{(2m+1)^2}  = 1 + \sum_{m=1}^{\infty}
\frac{a_{2m}}{(2m+1)^2}=
\nonumber \\
& &  
\! \! \! \! \! 
= 1 + \sum_{m=1}^{\infty}\frac{1}{(2m+1)^2}  
\frac{1}{2\pi i} {\oint}_{0} dz
\frac{1}{z^{2m+1}}\left(\frac{1+z}{1-z}\right)^{1/3} =
\nonumber \\
& &
\! \! \! \! \! \! \! \! \! \!
= 1+
\frac{\sqrt{3}}{2\pi}\int_{1}^{\infty} \! \!\! \!
dx\int_{0}^{1}\frac{dy}{y}\left[
\frac{1}{2} 
\ln \left(\frac{x+y}{x-y}\right)-\frac{y}{x}\right]
\left[\left(\frac{x+1}{x-1}\right)^{1/3}-
\left(\frac{x+1}{x-1}\right)^{-1/3}\right],
\nonumber 
\end{eqnarray}                                 
making the change of variables $(x-1)/(x+1) \rightarrow x $ 
the above expression reduces to 
\begin{eqnarray}
&& 
1+\frac{\sqrt{3}}{\pi}\int_{0}^{1} dx \left(x^{1/3}-x^{-1/3}\right)\left\{
\frac{1}{(1-x)(1+x)}-\frac{1}{2}\frac{1}{(1-x)^{2}}\right.
\nonumber \\
&&
\left. \int_{0}^{1}\frac{dy}{y}\ln
\left(\frac{\alpha(y)+x\gamma(y)}{\gamma(y)+x\alpha(y)}\right)
(x^{1/3}-x^{-1/3})\right\} \, ,
\nonumber
\end{eqnarray}
where $\alpha(y)=1+y$ and $\gamma(y)=1-y$. Similarly one rewrite all the sums
in integral forms. Converting all the sums in the expression for $\Delta_{1}$
to integrals and making use of the fact, 
$$
\frac{1}{(a+bx)^{2}} = -\frac{1}{b}\left(\frac{1}{a+bx}\right)^{\prime}\,\,,
$$
the expression for $\Delta_{1}$, after performing several integrations
(when integrating by parts it is crucial that one takes the
limits at the end of the calculation to preserve all the divergent pieces; they
do cancel among themselves rendering the whole quantity
finite), reduces to
$$
\Delta_{1} = 3- \frac{4}{\pi^{2}} \left[ \psi^{\prime} (4/3) - \beta^{\prime}
(4/3) - \beta^{\prime} (2/3) + \psi^{\prime} (2/3) \right] = \frac{1}{3} \, .
$$     

Now the value of
$\Delta_{1}$ is used to calculate the value of $ {(E(U'+
{\overline{U}}^{\prime})E)}_{11}$
which is needed as a boundary condition for
${(E(U'+{\overline{U}}^{\prime})E)}_{n\,\,n = odd}$. Skipping the details, we 
get 
\begin{eqnarray}
{\left( E(U'+{\overline{U}}^{\prime})E \right)}_{n\,\,m} & = &
2(-)^{n+1} \left( \frac{A_{n} B_{m} + B_{n} A_{m}}{n+m} + 
\frac{A_{n} B_{m} - B_{n} A_{m}}{n-m} \right)
\nonumber \\
& + &
\left(\frac{2}{n}\right)^{1/2} \left(\frac{2}{m}\right)^{1/2} \delta_{nm}
\, ,
\nonumber
\end{eqnarray}
for $n =odd $, $m = odd $ which is the case of interest (for 
$ n = even $, $ m = even $ it is more convenient to use the expressions given 
by (\ref{A3}) and (\ref{A7})). It is clear that for the off diagonal elements 
the above expression reduces to that in (\ref{A3}). For the diagonal 
elements it is understood that one must consider a limiting procedure 
( $ m \rightarrow n $ ). If
one takes the limit ($m\rightarrow n$) one recovers the expression in
(\ref{A7}) for $ n = odd $. Below we shall give the relevant limit needed to
recover (\ref{A7}) for $ n = odd$. Consider                                
$$
\lim_{n\rightarrow m} \frac{a_{2n-1} b_{2m-1} - b_{2n-1} a_{2m-1}}{(2n-1)
-(2m-1)}  \, .
$$
Writing the above expression in  integral representation we get
\begin{eqnarray}
&  \lim_{\epsilon\rightarrow {0}} &
\left(\frac{sin{\pi/3}}{\pi}\right)^{2}
\frac{1}{2\epsilon} \int_{1}^{\infty} \int_{1}^{\infty} dx dy
\left( \frac{1}{x^{2m+2\epsilon}} \frac{1}{y^{2m}} 
-\frac{1}{x^{2m}} \frac{1}{y^{2m+2\epsilon}}\right)  
\nonumber \\
& &                                
\left[ \left(\frac{x+1}{x-1}\right)^{1/3} +
\left(\frac{x+1}{x-1}\right)^{-1/3} \right]   
\left[ \left(\frac{y+1}{x-1}\right)^{2/3} +
\left(\frac{y+1}{y-1}\right)^{-2/3}\right]  \, .
\nonumber
\end{eqnarray}
Using the fact
$$
\lim_{\epsilon\rightarrow {0}}  
\frac{x^{-2\epsilon}-y^{-2\epsilon}}{2\epsilon} 
 =  ln \left(\frac{y}{x}\right) \, ,
$$
the above expression reduced to
\begin{eqnarray}           
& - &\! \!\! \!\! \!\! \!
\frac{\sqrt{3}}{2\pi} b_{2m-1} \int_{1}^{\infty}
\frac{dx}{x^{2m}} lnx  \left[\left(\frac{x+1}{x-1}\right)^{1/3} +
\left(\frac{x+1}{x-1}\right)^{-1/3} \right]  
\nonumber \\           
& + & \frac{\sqrt{3}}{2\pi} a_{2m-1}\int_{1}^{\infty}
\frac{dx}{x^{2m}}lnx  \left[\left(\frac{x+1}{x-1}\right)^{2/3} +
\left(\frac{x+1}{x-1}\right)^{-2/3} \right] =
\nonumber \\
& = & \! \!\! \!\! \!\! \!
- \frac{\sqrt{3}}{\pi} b_{2m-1}\sum_{n=0}^{\infty} a_{2n}
\int_{1}^{\infty} \frac{dx}{x^{2(m+n)}}lnx +
\frac{\sqrt{3}}{\pi} a_{2m-1}\sum_{n=0}^{\infty} b_{2n}
\int_{1}^{\infty} \frac{dx}{x^{2(m+n)}}lnx  =
\nonumber \\
& = & 
\! \!\! \!\! \!\! \!
- \frac{\sqrt{3}}{\pi}\left[a_{2m-1}{\tilde{\Sigma}}_{2m-1}^{b}-
b_{2m-1}{\tilde{\Sigma}}_{2m-1}^{a} \right] \, .
\nonumber \\
\label{A21}
\end{eqnarray}
A particular case of interest is the value of the limit when $m = 1$. In this
case, equation (\ref{A21}) gives
\be
\lim_{n\rightarrow 0}  \frac{  b_{1} a_{2n-1} - a_{1} b_{2n-1}}{2n} 
=\frac{\sqrt{3}}{\pi}\left[a_{1}{\tilde{\Sigma}}_{1}^{b}-
b_{1}{\tilde{\Sigma}}_{1}^{a} \right]= -\frac{1}{3}  \, .
\label{A22}
\end{equation}  
To see this, let us first consider ${\tilde{\Sigma}}_{1}^{a}$. Writing 
${\tilde{\Sigma}}_{1}^{a}$ in integral form we obtain
\be
{\tilde{\Sigma}}_{1}^{a} = -\frac{1}{2}\frac{d}{dk} I(k,q)|_{k=0,q=1/3}
\, ,
\label{A23}
\ee
where
$$
I(k,q) = \int_{0}^{1} dt t^{k} \left(\frac{1+t}{1-t}\right)^{q} 
       + \int_{0}^{1} dt t^{k} \left(\frac{1+t}{1-t}\right)^{-q} \, .
$$
This integral is easily evaluated using the $hypergeometric$ function
\be
F(a,b;c;z) = \frac{\Gamma (c)}{\Gamma (b) \Gamma (c-b)} \int_{0}^{1} dt
t^{b-1} (1-t)^{c-b-1} (1-zt)^{-a} \, .
\label{A25}
\ee
Using (\ref{A25}), and after some algebraic  manipulation of the the gamma
functions, equation (\ref{A23}) gives
\be
{\tilde{\Sigma}}_{1}^{a} = \frac{\pi}{\sqrt{3}}\left(ln\frac{3}{2} +
\frac{1}{6}\right) \, . 
\label{A26}
\ee

Similar steps give
\be  
{\tilde{\Sigma}}_{1}^{b} = \frac{2\pi}{\sqrt{3}}\left(ln\frac{3}{2} -
\frac{1}{12}\right) \, . 
\label{A27}
\ee
Substituting (\ref{A26}) and (\ref{A27}) in the LHS of (\ref{A22}) we obtain
the desired result.

\appendice 

\noindent
\underline{Sums Involving $a_{n}$ and $b_{n}$ Continue }.

A par\-ti\-cu\-lar 
com\-bi\-na\-tion of sums which appears in the text is
$$                                   
a_{1}\sum_{n=1}^{\infty} \frac{b_{2n+1}}{1-(2n+1)} -
b_{1}\sum_{n=1}^{\infty} \frac{a_{2n+1}}{1-(2n+1)} \, .
$$  
Which, when written in integral form, becomes
$$  
-\frac{2}{3}\sum_{n=1}^{\infty}\frac{1}{2n} \frac{1}{2\pi
i}\oint_{0}\frac{dz}{z^{2n+2}} \left[\left(\frac{1+z}{1-z}\right)^{2/3}-
2 \left(\frac{1+z}{1-z}\right)^{1/3}\right] \, .
$$  
Deforming the contour and picking up the integrals along the cuts we obtain

\begin{eqnarray}
& & \frac{1}{\pi\sqrt{3}}\int_{1}^{\infty}\frac{dx}{x^2}
\ln\left(\frac{x^2}{x^2-1}\right)\left[  \left(\frac{x+1}{x-1}\right)^{2/3} +
\left(\frac{x+1}{x-1}\right)^{-2/3} - 
\right.
\nonumber \\ 
& & \left.
- 2  \left(\frac{x+1}{x-1}\right)^{1/3}-
2 \left(\frac{x+1}{x-1}\right)^{-1/3}\right].
\nonumber 
\end{eqnarray}
Employing the change of variables
$(x-1)/(x+1) \rightarrow x $ the whole expression simplifies
$$
-\frac{1}{\pi\sqrt{3}}\int_{0}^{1}\frac{dx}{{(x+1)}^2}
\ln\left[\frac{{(x+1)}^2}{4x}\right]\left(x^{2/3} + x^{-2/3} - 2 x^{1/3} - 2
x^{-1/3}\right)  \, .
$$
This expression can be further reduced by integration by parts. Again one 
has to be careful since integration by parts give 
rise to many divergent terms as one approaches the branch points. These 
divergent terms have different magnitudes and one must 
keep a track of them 
as they conspire at the end to give a finite number. All in all, we end up
with
\be
-\frac{1}{\pi\sqrt{3}}\left[\frac{9}{2} - \beta(1/3) + 2\beta(2/3) -2\beta(4/3)
+\beta(5/3)\right] = -\frac{2}{3}.
\label{B05}
\ee

\noindent
\underline{Deriving The Various identities Used in The Text}.

First lets derive the identity in (\ref{CV26}). Consider 
$$
W_{mn} = W_{nm} \equiv \frac{a_m b_n + b_m a_n}{m+n} \, .
$$

Now it is not hard to see (by direct substitution) that 
\be
(m+1) W_{m+1\,n} - (m-1) W_{m-1\,n} + (n+1) W_{m\,n+1} -(n-1) W_{m\,n-1} = 0
\, ,
\label{B2}
\ee
for $ m+n$ = odd integer. Letting $m \rightarrow 2m-1$, $n \rightarrow 2n$ 
and summing over $m\geq 1$ we get
\be
(2n+1) W_{2n+1} = (2n-1) W_{2n-1} \, ,
\label{B3}
\ee
where
$$
W_{n = odd} = \sum_{m=1}^{\infty} \frac{a_{n } b_{2m-1} + b_{n} a_{2m-1}}
{(2m-1) + n}.
$$
From the recursion relation (\ref{B3}) it follows that
$$
W_{2n-1} = \frac{W_{1}}{2n-1} \, ,
$$
where
$$
W_{1} = \sum_{m=1}^{\infty} \frac{a_{1} b_{2m-1} + b_{1} a_{2m-1}}{2m}
\, .
$$

The value of $W_{1}$ can be evaluated using the integral representation of the
Taylor  coefficients. The steps involved in calculating $W_{1}$ are similar to
the ones leading to (\ref{B05}). Doing so we get $W_{1} = 2$ and  identity 
(\ref{CV26}) follows. Next consider the identity given in (\ref{CV26}).  If we
define
$$
T_{nm}=T_{mn}\equiv \frac{a_{m} b_{n} - b_{m} a_{n}}{m-n} \, ,
$$
then it follows that
$$
(n+1) T_{n+1\,m} - (n-1) T_{n-1\,m} + (m+1) T_{n\,m+1} - (m-1) T_{n\,m-1} = 0
\, , 
$$
for $m+n=$ odd integer.  Now letting $m\rightarrow 2m-1\geq 1$, 
$n\rightarrow 2n\geq 2$ and summing over $m$ we arrive at
\be
(2n+1) \sum_{m=1}^{\infty} T_{2n+1\,2m-1} = (2n-1) \sum_{m=1}^{\infty}
 T_{2n-1\,2m-1}
\, ,
\label{B9}
\ee
from which (\ref{CV26}) follows.

To derive (\ref{C9.1}) let $n \rightleftharpoons m $  in (\ref{B2}) and then 
letting  $n\rightarrow {2n-1}\geq 1$, $m\rightarrow {2m}\geq 2$ and summing 
over $m$ we get
\begin{eqnarray}
&& 2n \left(a_{2n} S_{2n}^{b} + b_{2n} S_{2n}^{a} \right) 
 -  (2n -2) \left(a_{2n-2} S_{2n-2}^{b} + b_{2n-2} S_{2n-2}^{a} \right)
\nonumber \\
 & = & \frac{2}{2n}\left[ a_{2n-1} + b_{2n-1} - a_{2n-2} b_{2n-2} \right]
\nonumber
\end{eqnarray}
The expression inside the square bracket is most easily evaluated using the 
integral representation of the Taylor coefficients. The resulting integrals 
are easily evaluated and yield zero. Hence, (\ref{B9}) reduces to 
\be
2n \left(a_{2n} S_{2n}^{b} + b_{2n} S_{2n}^{a} \right)  
-(2n -2) \left(a_{2n-2} S_{2n-2}^{b} + b_{2n-2} S_{2n-2}^{a} \right) = 0
\, .
\label{B10}
\ee
Solving (\ref{B10}) we get
\be
C_{2n} \equiv a_{2n} S_{2n}^{b} + b_{2n} S_{2n}^{a} = \frac{2}{2n} C_{2}
\, .
\label{B11}
\ee

This solution depends on the value of $C_{2}$ which is 
given by
$$
C_{2} = a_{2} S_{2}^{b} + b_{2} S_{2}^{a} = 1 \, .
$$
It can be easily found using the results:
$$
S_{2}^{a(b)} = \frac{1}{2} a(b) S_{1}^{a(b)} +\frac{1}{2} \, 
$$
and (\ref{CV26}). Substituting the value of $C_{2}$ back into (\ref{B11}) we
arrive at (\ref{C9.1}).

We conclude this Appendix by deriving two more identities which are needed in
the proof of the $momentum$ overlaps.  Consider the recursion relation given 
in (\ref{TS18})

\be
(n+1) {\tilde{\Sigma}}_{n+1}^{a} = \frac{2}{3} {\tilde{\Sigma}}_{n}^{a}
+(n-1) {\tilde{\Sigma}}_{n-1}^{a} + \Sigma_{n+1}^{a} -\Sigma_{n-1}^{a}
\, ,
\label{B18}
\ee                                                        
Letting $n\rightarrow {-n}$ in the above expression and subtracting the result
from $\frac{1}{2} \times $ equation (\ref{B18}) and remembering that
$\Sigma_{-n}^{a}=-\frac{1}{2}\Sigma_{n}^{a}$, we get
\begin{eqnarray}
&&
(n+1)\left[\frac{1}{2}{\tilde{\Sigma}}_{n+1}^{a}-{\tilde{\Sigma}}_{-(n+1)}^{a}
\right]
-(n-1)\left[\frac{1}{2}{\tilde{\Sigma}}_{n-1}^{a}-{\tilde{\Sigma}}_{-(n-1)}^{a}
\right]
\nonumber \\
&&
- \frac{2}{3}\left[\frac{1}{2}{\tilde{\Sigma}}_{n}^{a}-{\tilde{\Sigma}}_{-n}^{a}
\right] = 0 \, .
\label{B19}
\end{eqnarray}
The above equation is the same as (\ref{TS16}) and therefore has a solution
proportional to $S_{n}^{a}$. Hence\footnote{From the integral representation of
$S_{n}$, it follows that when $n\rightarrow {0} $, $S_{n}\rightarrow {c/n}$ as
required.} 
\be
\frac{1}{2}{\tilde{\Sigma}}_{n}^{a}-{\tilde{\Sigma}}_{-n}^{a} =\kappa S_{n}^{a}
\label{B20}
\ee
This solution depends on the constant $\kappa$ that we obtain
setting $n=1$ in (\ref{B20}) 
$$
\kappa = \frac{1}{S_{1}^{a}}\left(\frac{1}{2}{\tilde{\Sigma}}_{1}^{a}-
{\tilde{\Sigma}}_{-1}^{a}\right)  \, ,
$$
and direct substitution yields
$$
\kappa = - \frac{3}{2} \Sigma_{0}^{a} \, .
$$
Substituting this into (\ref{B20}) and rearranging terms we obtain
$$
{\tilde{\Sigma}}_{-n}^{a} =
 \frac{1}{2}{\tilde{\Sigma}}_{n}^{a} + \frac{3}{2}\Sigma_{0}^{a} S_{n}^{a}
\, .
$$

Similarly starting with(\ref{TS18}) we can easily obtain
${\tilde{\Sigma}}_{n}^{b}$ for negative values of $n$ 
$$
{\tilde{\Sigma}}_{-n}^{b} =
- \frac{1}{2}{\tilde{\Sigma}}_{n}^{b} + \frac{1}{2}\Sigma_{0}^{b} S_{n}^{b}
\, .
$$

\end{document}